\documentclass[preprint,12pt]{elsarticle}
\usepackage{}

\usepackage{amsmath}
\usepackage{cases}
\usepackage{color}
\usepackage{hyperref}
\usepackage{amssymb}
\usepackage{graphicx}
\usepackage{graphics}
\usepackage{subfigure}
\usepackage{slashbox}
\usepackage{appendix}
\usepackage{textcomp}
\usepackage{bm}
\usepackage{url}

\makeatletter
  \newcommand\figcaption{\def\@captype{figure}\caption}
  \newcommand\tabcaption{\def\@captype{table}\caption}
\makeatother

\biboptions{numbers,sort&compress}

\begin{document}

\begin{center}
{\bf \large HAM-Schr\"{o}dingerisation: a generic framework of quantum simulation for any nonlinear PDEs} \\ \vspace{0.5cm}

Shijun Liao$^{1,2}$ \footnote{The corresponding author,  email address: sjliao@sjtu.edu.cn}  \\ \vspace{0.3cm}

$^{1}$ State Key Laboratory of Ocean Engineering, Shanghai 200240, China \\
$^{2}$ School of Ocean and Civil Engineering, Shanghai Jiaotong University,  Shanghai 200240, China \\

\end{center}

\hspace{-0.75cm} {\bf Abstract}
{ 
Recently, Jin {\em et al.} proposed a quantum simulation technique for  {\bf any}  {\em linear} partial differential equations (PDEs), called  Schr\"{o}dingerisation~\cite{jin2022arXivA, jin2022arXivB, Jin2023PRA}.  In this paper,   the Schr\"{o}dingerisation technique for quantum simulation is expanded to  {\bf any}   {\em nonlinear} PDEs by combining it with the homotopy analysis method (HAM) \cite{Liao1992PhD, Liao2003Book, Liao2012Book}.  The HAM  can transfer a nonlinear PDE into a series of linear PDEs with guaranteeing  convergence of the series.   In this way,  {\bf any} nonlinear PDEs  can be solved by quantum simulation  using a quantum computer.  For simplicity, we call  the procedure  ``HAM-Schr\"{o}dingerisation quantum algorithm''.   Quantum computing is a groundbreaking technique.  Hopefully, the  ``HAM-Schr\"{o}dingerisation quantum algorithm'' can open a door to highly efficient simulation  of complicated turbulent flows  by  means  of  quantum  computing  in future.            
}   

\section{Introduction} 

Today, quantum computing \cite{Harrow2009PRL, Berry2014JPAMT, Childs2017SIAM, Berry2017CMP, Subasi2019PRL, Childs2020CMP, Costa2021arXiv, YangYue2024PCI, YangYu2024CP} offers a core opportunity for computational methods.   Hamiltonian simulation is likely to be of particular importance in quantum computing, and is valid for the following time-dependent Schr\"{o}dinger equation   
\begin{equation}
i \; \partial_t \psi = H(t) \; \psi, 
\end{equation}
where $H(t)$ is a time-dependent Hamiltonian operator, $\psi$ is a function,  $t$ denotes the time,  and  $i=\sqrt{-1}$, respectively.    

Jin et al \cite{jin2022arXivA, jin2022arXivB, Jin2023PRA} introduced a generic framework,  called  Schr\"{o}dingerisation, which  can map  {\em any} linear PDEs into Schr\"{o}dinger equations in real time.   This is  a  milestone in quantum simulation.   Based on a new approach  called  ``warped phase transformation'',  Schr\"{o}dingerisation can be used to solve {\em any} system of linear PDEs using quantum simulation, where the general form of the PDEs is given by
\begin{equation}
\frac{\partial\psi({\bf r},t)}{\partial t} = {\cal L}[\psi({\bf r},t)] + f({\bf r},t),   \;\;\; {\bf r} \in \Omega, t\geq 0,  \label{Linear-PDE-geq}
\end{equation}
subject to the initial  condition
\begin{equation}
\psi({\bf r},0) =\alpha({\bf r})  \label{Linear-PDE-ic}
\end{equation}
and the boundary condition
\begin{equation}
\psi({\bf r},t) = \beta(t) \;\; \mbox{when ${\bf r} \in \Gamma$},  \label{Linear-PDE-bc}
\end{equation}
where $\bf r$ and $t$ denote the spatial and temporal variable,  $\psi({\bf r},t)$ is a unknown function, $\cal L$ is a linear operator, $f({\bf r},t), \alpha({\bf r})$, and $\beta(t)$ are known functions, $\Omega$  denotes the physical domain and $\Gamma$ denotes its boundary, respectively.   In practice, a linear PDE in the form (\ref{Linear-PDE-geq}) - (\ref{Linear-PDE-bc})  can be discretized  in space to get a system of linear ordinary differential equations (ODEs) as follows:
\begin{equation}
\frac{d {\bf u}(t)}{d t} = A(t) {\bf u}(t) + {\bf b}(t),  \;\; {\bf u}(0) = {\bf a},  \label{linear-ODE}
\end{equation}
where ${\bf b} \in \mathbb{C}^n$ and  ${\bf u} \in \mathbb{C}^n$  are known functions,  ${\bf a} \in \mathbb{C}^n$ is a known vector,  $A(t) \in \mathbb{C}^{n \times n}$ can be a non-Hermitian matrix, i.e.  $A(t)$ might not be equal to its conjugate transpose, respectively.   The key point is that the linear equation (\ref{linear-ODE}) can be solved using the quantum simulation technique   Schr\"{o}dingerisation  \cite{jin2022arXivA, jin2022arXivB, Jin2023PRA}.    Several applications are described in the literature 
illustrating the validity of Schr\"{o}dingerisation in solving many types of linear PDEs~\cite{Jin2024JCP, Jin2024QST, Jin2023arXivA, Jin2023arXivB, Jin2023arXivC, Jin2023arXivD, Jin2024arXivA, Jin2024arXivB, Jin2024arXivC, Hu2024arXiv}.  

An important question follows.  Can any {\em nonlinear} PDEs  be solved by means of quantum computing?     The answer is yes, as shown in Section~2.        

\section{HAM-based quantum simulation for nonlinear PDEs}

More specifically, can any nonlinear PDE be transferred into a series of linear PDEs with {\em convergence guarantee} of the solution series?  The answer is yes, as described below.    

Mechanics, as a significant branch of natural sciences, often deals with the core problem of solving nonlinear equations. To do a good job, one must first sharpen one's tools.  Continuously breaking through the limitations of traditional methods and proposing more effective new approaches is one of the essential tasks in modern mechanics.  Analytical solutions possess unique advantages, as they can uncover the universal laws and essential characteristics of a problem. Traditional analytical approximation methods, represented by perturbation methods, typically rely on small physical parameters and often encounter issues such as solution divergence or slow convergence, thus are generally only applicable to weakly nonlinear problems.  

In 1992 Liao \cite{Liao1992PhD} proposed a new analytic approximation method for highly nonlinear equations (including ODEs, PDEs and all other types of nonlinear equations), namely the homotopy analysis method (HAM) \cite{Liao2003Book, Liao2012Book, Liao1997IJNLM, Liao1999IJNLM, Liao2004AMC, Liao2007SAM, Liao2009CNSNS, Liao2010CNSNS, Liao2016NA, Liao2016JFM, Zhong2018JFM, Liao2020SCPMA, Liao2023AAMM}.  Homotopy is a basic concept in topology that describes continuous deformations.  However,  using this classical concept of homotopy, one still  cannot guarantee the convergence of solution series.  So, Liao \cite{Liao1999IJNLM}  proposed a totally new concept, namely  the ``generalized   homotopy'',   by means of introducing a new auxiliary parameter $c_{0}$, called ``convergence-control parameter'' that has no physical meanings,  to greatly generalize the classical concept of homotopy.   Unlike traditional methods,  the HAM can transfer any  nonlinear equation into a series of linear sub-equations without any physical assumptions:   the solution of the original nonlinear equations is equal to the summation of the solutions of these linear sub-equations.  Firstly, unlike perturbation methods and all other approximation techniques for nonlinear equations, the convergence of the solution series given by the HAM is guaranteed by means of the so-called convergence-control parameter $c_0$ \cite{Liao1999IJNLM, Liao2010CNSNS}.  
For example, one cannot gain convergent results for limiting Stokes waves in extremely shallow water by means of perturbation
methods, even with the aid of extrapolation techniques such as the Pad\'{e} approximant.  In particular, it is extremely difficult for traditional analytic/numerical approaches to present the wave profile of limiting waves with a sharp crest of 120$^{\circ}$  included angle
first mentioned by Stokes in the 1880s.  However, using the HAM, we successfully gain convergent results (and especially the wave profiles) of the limiting Stokes waves with this kind of sharp crest in arbitrary water depth,  even including solitary waves of extreme form in extremely shallow water, without using any extrapolation techniques \cite{Zhong2018JFM}.     Secondly, unlike perturbation methods, the HAM provides the great freedom to choose the type of the linear sub-equations \cite{Liao2007SAM, Liao2016NA, Liao2016JFM, Liao2023AAMM} and also bestows the great freedom in the choice of initial guess solution. 
For example, the so-called ``small denominator problem''  is a fundamental problem of dynamics, which appears  most commonly in
perturbative theory.  However,  ``small denominator problem'' can be completely avoided  \cite{Liao2023AAMM}  by means of a non-perturbative approach based on HAM, namely ``the method of directly defining inverse mapping'' (MDDiM) \cite{Liao2016NA} which can solve a nonlinear equation by directly ``defining'' an inverse operator, say, without ``calculating''  its inverse operator, mainly because  the HAM provides great freedom to {\em choose} auxiliary linear operator.     
To date,  several thousand articles related to the HAM have been published in a wide range of fields including applied mathematics, physics, engineering, nonlinear mechanics,   quantum mechanics,  bio-mechanics,  astronomy, finance and so on  \cite{ Song2007, Nassar2011, Mastroberardino2011, Kimiaeifar2011CMA, Sardanyes2015, VanGorder2017, Pfeffer2017, KV2018NA, Cullen2019JCP, Sultana2019EPJP,  Botton2022AMM, Masjedi2022AMM, Kaur2022JMAA, Liu2021JMPS, Liu2023IJSS}.        

Without loss of generality, let  us consider the general nonlinear PDE
\begin{equation}
\frac{\partial\psi({\bf r},t)}{\partial t} = {\cal N}[\psi({\bf r},t)]  + g({\bf r},t), \;\; {\bf r}\in \Omega, t \geq 0, 
 \label{NL-PDE-geq}
\end{equation} 
subject to the initial  condition
\begin{equation}
\psi({\bf r},0) =\alpha({\bf r})  \label{NL-PDE-ic}
\end{equation}
and the boundary condition
\begin{equation}
\psi({\bf r},t) = \beta(t) \;\; \mbox{when ${\bf r} \in \Gamma$},  \label{NL-PDE-bc}
\end{equation}
where $\bf r$ and $t$ denote the spatial and temporal variable,  $\psi({\bf r},t)$ is a unknown function, $\cal N$ is a nonlinear operator, $g({\bf r},t), \alpha({\bf r}) $ and $\beta(t)$ are known functions, and $\Omega$ and $\Gamma$ denote the physical domain and its boundary, respectively. 

In the frame of the HAM, let $q \in[0,1]$ denote an embedding parameter that has no physical meaning,  $\psi_0({\bf r},t)$  be  a guess solution of  $\psi({\bf r},t)$,  the non-zero constant $c_0$ be the convergence-control parameter, and ${\mathcal L}^\star$ denote an auxiliary linear operator with the property ${\mathcal L}^\star[0] = 0$.   It should be emphasized here  that one has great freedom to choose the guess solution $\psi_0({\bf r},t)$  and the auxiliary linear operator  ${\cal L}^\star$.  Then, a family of solutions $\Psi({\bf r},t,q)$, where $q\in[0,1]$ is an embedding parameter in topology, is constructed by the following zeroth-order deformation equation
\begin{equation}
(1-q){\cal L}^\star\left[\Psi({\bf r},t,q) - \psi_0({\bf r},t) \right] = c_0 \; q \;\left\{ \frac{\partial\psi({\bf r},t)}{\partial t} - {\cal N}[\psi({\bf r},t)]  - g({\bf r},t)\right\}, \label{geq-zero}
\end{equation} 
subject to the initial  condition  
\begin{equation}
\Psi({\bf r},0,q)  = (1-q)\; \psi_0({\bf r},0)  + q \; \alpha({\bf r}) \hspace{0.75cm} \mbox{when $t=0$},  \label{ic-zero}
\end{equation}
and the boundary condition 
\begin{equation}
\Psi({\bf r},t,q) =(1-q)\; \psi_0({\bf r},t)  + q\;  \beta(t) \hspace{0.75cm}  \mbox{when ${\bf r} \in \Gamma$}.  \label{bc-zero}
\end{equation}
Due to the property  ${\cal L}^\star[0] = 0$, the solution of Eqs. (\ref{geq-zero}) - (\ref{bc-zero}) when $q=0$ is exactly  the initial guess, i.e.   
\begin{equation}
\Psi({\bf r},t,0)  = \psi_0({\bf r},t).   \label{Psi:q=0}
\end{equation}
When $q=1$,  Eqs. (\ref{geq-zero}) - (\ref{bc-zero})  are exactly the same as the original nonlinear PDEs (\ref{NL-PDE-geq})-(\ref{NL-PDE-bc}), thus 
\begin{equation}
\Psi({\bf r},t,1)  = \psi({\bf r},t).   \label{Psi:q=1}
\end{equation}
So, the solution $\Psi({\bf r},t,q)$ of Eqs. (\ref{geq-zero}) - (\ref{bc-zero}), where $q\in[0,1]$ is the embedding parameter, connects the known guess solution $\psi_0({\bf r},t)$ and the unknown solution $\psi({\bf r},t)$ of the original nonlinear PDEs (\ref{NL-PDE-geq})-(\ref{NL-PDE-bc}).  Since one has great freedom  \cite{Liao2007SAM, Liao2016NA, Liao2023AAMM} to choose the guess solution  $\psi_0({\bf r},t)$ and the auxiliary linear operator ${\cal L}^\star$, it is reasonable to assume that both of  $\psi_0({\bf r},t)$ and ${\cal L}^\star$ are so properly chosen that  $\Psi({\bf r},t,q)$ 
exhibits continuous deformation with respect to the embedding parameter $q\in[0,1]$, say, as $q$ increases from 0 to 1,  $\Psi({\bf r},t,q)$ deforms {\em continuously} from the known guess solution $\psi_0({\bf r},t)$ to the unknown solution $\psi({\bf r},t)$ of the original nonlinear PDEs (\ref{NL-PDE-geq})-(\ref{NL-PDE-bc}),  and moreover  $\Psi({\bf r},t,q)$ can be expanded into a Taylor series with respect to $q$ at $q=0$, say,
\begin{equation}
\Psi({\bf r}, t, q) =  \psi_0({\bf r},t) + \sum_{m=1}^{+\infty} \psi_m({\bf r},t) \; q^m,  \label{Psi:Taylor-series}
\end{equation}
where 
\begin{equation}
\psi_m({\bf r},t)  = \left. \frac{1}{m!} \frac{\partial^m \Psi({\bf r}, t, q)}{\partial q^m}\right|_{q=0}.
\end{equation}
 Note that a Taylor series often has a finite  radius of convergence so that the Taylor series (\ref{Psi:Taylor-series}) might be divergent at $q=1$.   Fortunately, in the HAM framework, one also has great freedom in the choice of  the convergence-control parameter $c_0$.   Assuming that the guess solution $\psi_0({\bf r},t)$,  the auxiliary linear operator ${\cal L}^\star$ and the  convergence-control parameter $c_0$ are selected such that the Taylor series (\ref{Psi:Taylor-series}) is convergent at $q=1$, one has due to (\ref{Psi:q=1}) the homotopy series solution    
\begin{equation}
\psi({\bf r}, t) =  \psi_0({\bf r},t) + \sum_{m=1}^{+\infty} \psi_m({\bf r},t).  \label{psi:series-solution}
\end{equation}
In practice, only finite terms are used. So, one has the $M$th-order homotopy approximation
\begin{equation}
\psi({\bf r}, t)  \approx   \psi_0({\bf r},t) + \sum_{m=1}^{M} \psi_m({\bf r},t).  \label{psi:series-solution:Mth}
\end{equation}
Note that the HAM provides us the great freedom in the choice of guess solution $ \psi_0({\bf r},t)$.  Logically, the $M$th-order homotopy approximation should be a better approximation than the guess solution $ \psi_0({\bf r},t)$, so long as the convergence-control parameter $c_0$ is properly chosen.  So, one can use the $M$th-order homotopy approximation as a new guess solution to  further gain a better $M$th-order homotopy approximation.  This provides us a $M$th-order iterative formula, which works quite well even at low order \cite{Liao2003Book, Liao2012Book}.    Here the key point is that $\psi_m({\bf r},t)$ in (\ref{psi:series-solution}) is governed by a {\em linear} PDE, as described below in detail.   Note that  artificial noise caused by quantum simulation can be decreased by iterations.   

Differentiating both sides of the zeroth-order deformation equations (\ref{geq-zero}) - (\ref{bc-zero}) $m$ times with respect to $q$ and then dividing them by $m!$ and finally setting $q=0$, one has the {\em linear} $m$th-order deformation equation ($m\geq 1$)
\begin{equation}
{\cal L}^\star\left[ \psi_m({\bf r},t) - \chi_m \;  \psi_{m-1}({\bf r},t)    \right] = c_0 \; \Delta_{m-1}({\bf r},t),  \label{mth-geq}
\end{equation}
subject to the initial  condition  
\begin{equation}
\psi_m({\bf r},0)  = \left\{   
\begin{array}{cl}
 \alpha({\bf r})-\psi_0({\bf r},0) & \mbox{when $m=1$}\\
 0 & \mbox{otherwise}
\end{array}
\right.  \label{mth-ic}
\end{equation}
and the boundary condition 
\begin{equation}
\psi_m  = \left\{   
\begin{array}{cl}
 \beta(t) -\psi_0({\bf r},t) & \mbox{when $m=1$ and ${\bf r} \in \Gamma$}\\
 0 & \mbox{otherwise}
\end{array}
\right.  \label{mth-bc}
\end{equation}
where 
\begin{equation}
\chi_m = \left\{ 
\begin{array}{ll}
0 & \mbox{when $m \leq 1$} \\
1 & \mbox{otherwise}
\end{array}
  \right.
\end{equation}
and 
\begin{eqnarray}
\Delta_0({\bf r},t) &=& \frac{\partial\psi_0({\bf r},t)}{\partial t} - {\cal N}[\psi_0({\bf r},t)]  - g({\bf r},t),\\
\Delta_k({\bf r},t) &=& \frac{\partial\psi_k({\bf r},t)}{\partial t} - \frac{1}{k!}\frac{\partial^k{\cal N}[\Psi({\bf r},t,q)] }{\partial q^k}, \hspace{0.75cm} k\geq 1.
\end{eqnarray}
Note that $\Delta_{m-1}({\bf r},t)$ is a function dependent upon $\psi_0({\bf r},t)$, $\psi_1({\bf r},t)$,$ \cdots$, $\psi_{m-1}({\bf r},t)$ and thus is {\em known} for the linear $m$th-order deformation equation.  

It should be emphasized that the $m$th-order deformation equations (\ref{mth-geq}) - (\ref{mth-bc})  are {\em linear}!  Noting again that the HAM permits  the great freedom in the choice of  auxiliary linear operator \cite{Liao2007SAM, Liao2016NA, Liao2016JFM, Liao2023AAMM},  we set 
\begin{equation}
{\cal L}^\star[ \psi({\bf r},t)] =  \frac{\partial\psi({\bf r},t)}{\partial t} - {\cal L}[\psi({\bf r},t)],
\end{equation}
where $\cal L$ is a linear operator, which we also have great freedom to choose.  Then,  the linear  $m$th-order deformation equation (\ref{mth-geq}) can be written as 
\begin{equation}
\frac{\partial\delta_m({\bf r},t)}{\partial t} = {\cal L}[\delta_m({\bf r},t)] + f_m({\bf r},t),   \label{PDE:geq:m}
\end{equation}
where 
\begin{equation}
f_m({\bf r},t) = c_0 \; \Delta_{m-1}({\bf r},t)  \label{def:f_m}
\end{equation}
is a known function and 
\begin{equation}
\delta_m({\bf r},t) =  \psi_m({\bf r},t) - \chi_m \;  \psi_{m-1}({\bf r},t),  \label{def:delta_m}
\end{equation}
which gives 
\begin{equation}
\psi_m({\bf r},t)   =  \delta_m({\bf r},t)  + \chi_m \;  \psi_{m-1}({\bf r},t).
\end{equation}

Note that the linear PDE  (\ref{PDE:geq:m}) is exactly the same as (\ref{Linear-PDE-geq}), and thus can be solved using quantum simulation by Schr\"{o}dingerisation \cite{jin2022arXivA, jin2022arXivB, Jin2023PRA}.  In this way, an approximate solution of the original nonlinear PDEs (\ref{NL-PDE-geq}) - (\ref{NL-PDE-bc}) can be obtained by quantum computing.   If  this quantum simulation result is not accurate enough, one can further use it as a new guess solution $\psi_0({\bf r},t)$ to gain a better approximation by quantum computing, and so on.   The key point here is that the HAM can {\em guarantee} the convergence of the solution series or the iteration approach by choosing a proper value for the convergence-control parameter $c_0$ as has been illustrated in several thousands of HAM publications \cite{Liao2003Book, Liao2012Book, Liao1997IJNLM, Liao1999IJNLM, Liao2004AMC, Liao2007SAM, Liao2009CNSNS, Liao2010CNSNS, Liao2016NA, Liao2016JFM, Zhong2018JFM, Liao2020SCPMA, Liao2023AAMM, Song2007, Nassar2011, Mastroberardino2011, Kimiaeifar2011CMA, Sardanyes2015, VanGorder2017, Pfeffer2017, KV2018NA, Cullen2019JCP, Sultana2019EPJP,  Botton2022AMM, Masjedi2022AMM, Kaur2022JMAA, Liu2021JMPS, Liu2023IJSS}. 

It should be emphasized that  the linear PDE  (\ref{PDE:geq:m}) can be solved in the frame of the HAM by means of the MDDiM, i.e.  method of directly defining inverse mapping \cite{Liao2016NA, Liao2023AAMM} , mainly because the HAM provides us great freedom to choose an auxiliary linear operator.  So, it would be great  if such kind of freedom can be combined with quantum computation.      

\section{Conclusions and discussions}  

Recently, a quantum simulation technique for any {\em linear} PDE, called  Schr\"{o}dingerisation \cite{jin2022arXivA, jin2022arXivB, Jin2023PRA}, is proposed by Jin~\cite{Jin2024JCP, Jin2024QST, Jin2023arXivA, Jin2023arXivB, Jin2023arXivC, Jin2023arXivD, Jin2024arXivA, Jin2024arXivB, Jin2024arXivC, Hu2024arXiv}.  In 1992, the so-called homotopy analysis method (HAM) was proposed by Liao~\cite{Liao1992PhD}.   Unlike perturbation methods and other techniques, the HAM can guarantee the convergence of solution series even in case of quite high nonlinearity, and moreover bestows us great freedom in choice of auxiliary linear operator and initial guess \cite{Liao2003Book, Liao2012Book}.   In this paper,  a generic framework of quantum computing for any nonlinear PDEs is described briefly by means of combining the HAM with the Schr\"{o}dingerisation technique, called the HAM-Schr\"{o}dingerisation quantum computing.  In this way, any nonlinear PDEs in generic form can be solved by quantum simulation using a quantum computer.  

Note that  quantum speedup of Schr\"{o}dingerisation technique has been illustrated by Jin~\cite{Jin2024JCP, Jin2024QST, Jin2023arXivA, Jin2023arXivB, Jin2023arXivC, Jin2023arXivD, Jin2024arXivA, Jin2024arXivB, Jin2024arXivC, Hu2024arXiv}  via various types of linear PDEs.  Especially,  Xue~et~al~\cite{xue2024arXiv} currently proposed a quantum technique in the frame of HAM, illustrating its validity and quantum speedup by means of several examples.  Thus, the HAM-Schr\"{o}dingerisation approach described in this paper should have quantum speedup, which will be illustrated in the near future.    

Note that it is rather time-consuming to solve PDEs related to turbulent flows by means of classical simulation algorithms such as direct numerical simulation (DNS) \cite{Orszag1970DNS} and the clean numerical simulation (CNS) \cite{Liao2023book, Qin2022JFM,  Qin2024JOES, Liao_Qin_2025, Liao_Qin_2025_NEC}.  Quantum computer is a pioneering technology and quantum simulation is a groundbreaking technique, although there is a long way to go.   Hopefully, the HAM-Schr\"{o}dingerisation quantum computing can open a new door to very efficiently simulate complicated turbulent flows by quantum computer someday in future.  

\vspace{0.5cm}          
 
\hspace{-0.75cm} {\bf Acknowledgements} {Thanks to the anonymous reviewers for their valuable suggestions and constructive comments.}

\vspace{0.5cm}
\hspace{-0.75cm} {\bf Author ORCID} {Shijun Liao, https://orcid.org/0000-0002-2372-9502}

\section*{References}

\bibliographystyle{elsarticle-num}

\bibliography{HAM-quantum.bib} 

\end{document}